\documentclass[a4paper]{article}
\usepackage[pdftex]{graphicx}
\usepackage{float}
\usepackage{amsmath}
\usepackage{amsfonts}
\usepackage{fullpage}
\newcommand{\abs}[1]{\left|#1\right|}
\usepackage[utf8]{inputenc}
\usepackage[T1]{fontenc}
\usepackage{hyperref}
\usepackage[english]{babel}

\newcommand{\matr}[1]{\underline{\underline{#1}}}
\renewcommand{\vec}[1]{\underline{#1}}

\DeclareMathOperator{\sign}{sign}

\makeatletter
\def\blfootnote{\xdef\@thefnmark{}\@footnotetext}
\makeatother

\pdfoutput=1

\begin{document}

\title{Partial reconstruction of the rotational motion of Philae spacecraft during its landing on comet 67P/Churyumov–Gerasimenko}

\author{Tamás Baranyai\thanks{Budapest University of Technology and Economics, Hungary}, András Balázs\thanks{Wigner Research Centre for Physics, Hungary}, Péter L. Várkonyi$^*$}

\maketitle

\abstract{This paper presents a partial reconstruction of the rotational dynamics of the Philae spacecraft upon landing on comet 67P/Churyumov–Gerasimenko as part of ESA's Rosetta mission. We analyze the motion and the events triggered by the failure to fix the spacecraft to the comet surface at the time of the first touchdown. Dynamic trajectories obtained by numerical simulation of a 7 degree-of-freedom mechanical model of the spacecraft are fitted to directions of incoming solar radiation inferred from in-situ measurements of the electric power provided by the solar panels. The results include a lower bound of the angular velocity of the lander immediately after its first touchdown. Our study also gives insight into the effect of the programmed turn-off of the stabilizing gyroscope after touchdown; the important dynamical consequences of a small collision during Philae's journey; and the probability that a similar landing scenario harms the operability of this type of spacecraft.}

\section{Introduction}

After a ten-year journey across the Solar System and many complicated manoeuvres, the Rosetta spacecraft \cite{glassmeier07} with the Philae lander \cite{bibring07a,bibring07b,biele08} attached to it smoothly approached a small celestial body of 2-4 km in diameter, comet 67P/Churyumov-Gerasimenko. The spacecraft executed additional fine manoeuvres to fly a multitude of low and high altitude orbits around the comet, mapping its shape and surface in detail never seen before, and continues to observe it for more than two years. The Rosetta spacecraft and Philae lander were equipped with scientific instruments that delivered a wealth of new knowledge about the comet, in addition to spectacular pictures. 

\begin{figure}[h]
\begin{center}
\includegraphics[scale=0.3]{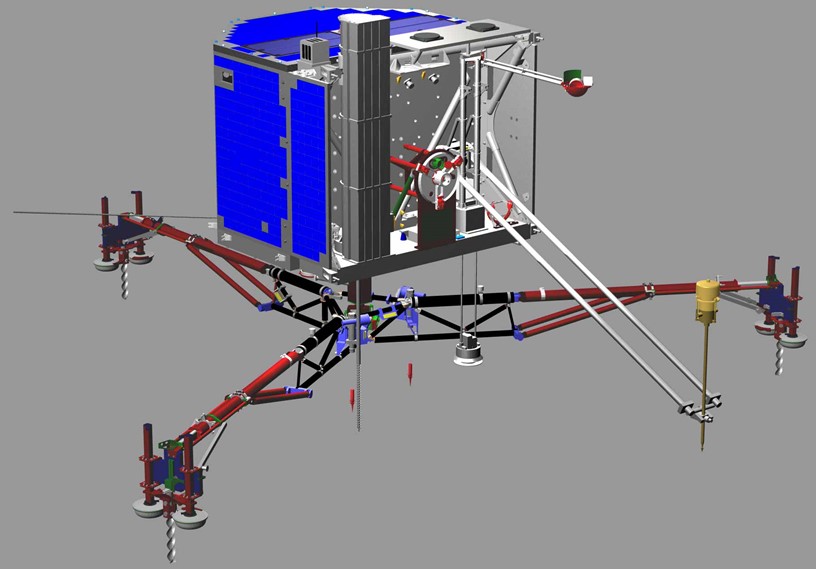}
\caption{Schematic structure of Philae (Image courtesy of DLR/Cologne)}
\label{fig:picture}
\end{center}
\end{figure}

The landing of Philae on the surface of the comet was initiated by the Rosetta spacecraft ~500 million km away from Earth, at a distance of 22.5 km from the comet on 12 November 2014. The lander reached the comet surface after a ballistic descent phase of ~7 hours, during which the attitude of the lander was partially stabilized by a single-axis gyroscope to ensure leg-forward landing on the surface (Fig. \ref{fig:picture}). The location and the attitude of the lander upon its first touchdown matched the target values, nevertheless the lander could not attach itself to the comet  due to an unexpected systematic failure in the dual redundant anchoring subsystem and a malfunction of the non-redundant active descent subsystem \cite{ulamec09,ulamec15,biele15}. 

The landing gear  \cite{witte14} absorbed enough energy to prevent complete escape from the weak gravitational field of the comet but the first touchdown was followed by several hours of uncontrolled tumbling motion until the spacecraft reached its final parking position roughly $1.25$ km away from the original target point. The lander remained functionally intact during its bouncing motion. It also kept radio contact alive and continuously sent all collected housekeeping and scientific telemetry data to the Rosetta spacecraft, which served as a relay station between Philae and Earth.

The rough topography of the final landing site strongly affected the energy supply of the lander
(via reduced exposure to solar radiation) as well as the reliability of the communication channel with
Rosetta, which prevented the Philae team from completing some parts of the planned scientic work.
Comunication with the lander was stopped ~60 hours after landing due to discharged batteries
of the lander. Afterwards, the lander fall into a state of hybernation for abot ~6 months, and an additional month was required to establish contact again with Philae. The exact location of the lander remains unknown \cite{herique15}.

Until now, most effort to understand what happened exactly during landing was devoted to the reconstruction of its translational motion motivated by the desire to locate its current position \cite{munoz15,biele15,esablog}. Much less work has been done to understand its rotational motion. Nevertheless we think that for planning future missions, it is highly desired to understand how the stabilizing gyroscope  affected the dynamics, whether or not solar panels may have hit the ground during tumbling motion (possibly resulting in damage) and how likely it was to come to rest with its legs on the ground, which was necessary for the operation of the lander. 

In the present work, we aim to reconstruct the rotational motion of the probe with the help of numerical simulation results, and in-situ measured data of the electric power produced by the solar power generators. This method takes advantage of having six solar panels covering Philae’s cube-like housing (Fig. \ref{fig:picture}). The electric power values of the six panels were sampled on a regular time basis with an approximate rate of 2 min/sample and delivered as telemetry data by the on-board computer of Philae \cite{balazs15}. A similar approach is described in \cite{topputo14} where the aim of the authors is to optimize lander attitude for maximum exposure to solar radiation.
%
We notice an ongoing parallel research with similar goals \cite{heinisch15}, in which measurements of the Rosetta Lander Magnetometer and Plasma Monitor (ROMAP) are used for attitude reconstruction, yet no results have been published yet.

The descent and the landing of Philae had four distinct phases, characterized by radically different types of motion. A schedule of events during landing is shown in Table \ref{tab:events}, see also \cite{biele15}. During the first descent phase (time: 08:35:00--15:34:04), slow, steady rotational motion occured without precession. The motion of the lander during this phase is simple and clearly understood. Solar power profile measurements during the descent phase serve as a verification of our method to reconstruct the direction of the radiation. A slight variation of the rotational motion took place at the time of unfolding the landing gear shortly after deparation from Rosetta, which will also be discussed briefly. The angular velocity changed abruptly upon the first touchdown (TD1) at time 15:34:04. Despite energy absorption of the landing gear, the failure of the anchoring devices allowed the initiation of fast rotational motion with a large precessional component. Precession and stability will be in the focus of the present paper. Another important effect was the automatic shutdown of the stabilizing gyroscope after TD1. The gyroscope gradually slowed down during the next 42 minutes due to internal friction, which triggered accelerating rotational motion of the lander itself because the conservation of angular momentum. A second abrupt change of rotational dynamics took place at 16:20 when the lander is suspected to hit a crater rim on the comet (event C1). There is limited amount of solar panel power data for the time after C1, and thus we rely primarily on the statistical analysis of simulation results with a variety of possible initial conditions to infer what may have happened with the spacecraft. Our results predict further increase of the precessional component, which dominated the rotational motion, implying that any part of the lander, including solar panels may have hit the comet surface at the time of the second touchdown at 17:25:26. After TD2, a short period of additional motion was followed by coming to rest at the final parking position in a rocky wedge, nevertheless with its legs touching the comet surface. 

The rest of the paper is structured as follows. In Section 2, we introduce the methods of the analysis, including the reconstruction of the direction of solar radiation from telemetry data, the mechanical model of the lander, and several distance metrics used for comparing measurements with simulation results. In Sec. 3, the main results of the analysis are presented. In particular, we verify the reconstruction method of Sun orbits (Sec. 3.1) using data of the descent phase; we estimate the angular velocity of the lander after TD1, with emphasis on initial values immediatey after TD1, and the amplitude of precession (Sec. 3.2); and finally, we analyse the possible effects of the collision at 16:20 (Sec. 3.3). The paper is closed by brief concluding remarks about the chosen strategy of landing and the presumptive effect of possible alternative control schemes.
\begin{table}[|c|c|]
\centering
\caption{Schedule of events during the landing of Philae}
\label{tab:events}
\begin{tabular}{|c|c|}
\hline
08:35:00 & separation from Rosetta               \\\hline
08:45-09:01        & gradual unfolding of landing gear \\\hline
15:34:04 & first touchdown; gyroscope turned off \\\hline
16:20:00 & collision with crater rim             \\\hline
17:25:26 & second touchdown                      \\\hline
17:31:17 & reaching final resting position      
\\ \hline
\end{tabular}
\end{table}

\section{Modelling methods}

\subsection{Reconstructing the directions of the Sun}

\begin{figure}[h]
\begin{center}
\includegraphics[scale=0.6]{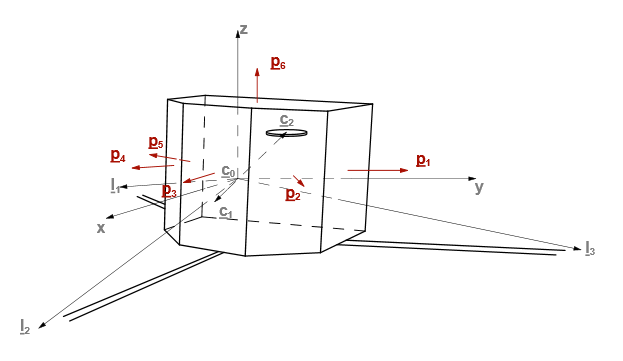}
\caption{Schematic view of Philae. Vectors $\vec{c}_0$, $\vec{c}_1$, $\vec{c}_2$ point to the center of mass of the whole lander, its main body and the gyroscope, respectively; $\vec{p}_i$ are normal vectors of the solar panels; $\vec{l}_i$ are vectors pointing to the endpoints of the legs.}
\label{fig:geometry}
\end{center}
\end{figure}

If solar panel $k$ is not shadowed by any obstacle, the electrical power produced by  the panel can be expressed as 
$$
P_k=\gamma_0\cdot C_k\cdot E_k\cdot D^{-2}\cdot \max(0,\vec{p}_k^T\vec{n})
$$
%
where $\gamma_0$ is a constant; symbol $^T$ denotes transpose; $C_k$ is the number of solar cells in panel $k$. In particular $C_1,...C_6$ are 254, 162, 162, 162, 254, 230, respectively. $E_k$ is the overall efficiency of panel $k$ and the converter linked to it.  $D$ is the distance of Sun from the lander; $\vec{p}_k$ is the unit normal vector of panel $k$ (Fig. \ref{fig:geometry}). Specifically,  $\vec{p}_1=[0,1,0]^T$, $\vec{p}_2=[2^{-1/2},2^{-1/2},0]^T$, $\vec{p}_3=[1,0,0]^T$, $\vec{p}_4=[2^{-1/2},-2^{-1/2},0]^T$, $\vec{p}_5=[0,-1,0]^T$, $\vec{p}_6=[0,0,1]^T$. $\vec{n}$ is a unit vector pointing towards the sun. The product $D^{-2}\max(0,\vec{p}_k^T\vec{n})$ determines the intensity of the incoming solar radiation. If $\vec{p}_k^T\vec{n}<0$, then radiation comes from behind the panel, i.e. it is shadowed. 

$E_k$ in general depends on incoming solar power and temperature. At relatively low solar powers at heliocentric distance of ~3 Astronomical Units (AU), where Philae's landing took place, these effects are negligible. Additionally, $E_k$ is influenced by the rate of degradation of the panel due to exposure to gamma radiation, nevertheless this has little effect during the relatively short time of the landing. Accordingly, we assume that $E_k$ is constant and equal for all $k$. Under these simplifying assumptions, we obtain
\begin{equation}
P_k=\gamma\cdot C_k\max(0,\vec{p}_k^T\vec{n}(t))
\label{eq: Pk(t)}
\end{equation}
where $\gamma$ is a constant. The exact value of $\gamma$ in ($\ref{eq: Pk(t)}$) depends sensitively on the charcteristics of components of the power subsystem, environmental parameters and the time history of the lander and thus we do not attempt to directly determine it. Instead, we define the normalized solar power distribution vector 
$$
\vec{P}(\vec{n})=\frac{[P_1\; P_2\; P_3\; P_4\; P_5\; P_6]^T}{||[P_1\; P_2\; P_3\; P_4\; P_5\; P_6]^T||}
$$
which does not depend on $\gamma$. The unknown vector $\vec{n}$ is reconstructed by numerical minimization of the error function
\begin{equation}
\Delta(\vec{n})=||\vec{P}(\vec{n})-\vec{P}_{measured}|| \label{eq:Delta}
\end{equation}
where $\vec{P}_{measured}$ is a solar power distribution vector obtained from in situ measurements.  Succesful reconstruction requires at least 2 independent constraints, for which at least 3 panels have to be illuminated. It is  also necessary that panel 6 is among the illuminated panels because $\vec{p}_1$, $\vec{p}_2$,...,$\vec{p}_5$ are all in the $x-y$ plane and thus the relative values of $P_1$,...,$P_5$ are independent of the $z$ coordinate of $\vec{n}$. Hence, succesful reconstruction of the Sun is not possible unless the Sun is 'above' the lander, i.e. $\vec{n}$ has a positive $z$ component. If there are more than 3 illuminated panels, then the data are overconstraining and $\Delta$ indicates the degree of self-inconsistency. Thus, $\Delta$ can be used to estimate the reliability of the reconstruction method.

Whenever the Sun is in a specific sector of space at lower altitudes 'behind’  the lander (in the direction of the negative x axis), a part of panel 6 becomes shadowed out by mechanical components mounted on the lander. This will result in a reduced value of $P_6$. The dependence of the output power on the number of shadowed cells is quite dramatic, which makes the reconstruction of the Sun direction vector in these ranges inaccurate. In order to avoid false reconstructions, we discarded all measurements in which panels 3 and 5 were simultaneously shadowed (which includes all directions prone to the shadowing effect).  

The results of the reconstruction are presented as a supplementary dataset. Relatively large $\Delta$ error values (in the range of 0.1-0.2) occur for data points measured before the deployment of the legs and for data points measured after coming to rest. These error values are attributed to unmodelled shadowing effects as discussed in Sec. 5. For the rest of the data points, the mean value of $\Delta(\vec{n})$ is 0.015, and the highest value is 0.07, which indicates a fairly reliable reconstruction. More tests of reliability will be conducted in Sec. 3.1. 

\begin{figure}[h!]
\begin{center}
\includegraphics[height=20 cm]{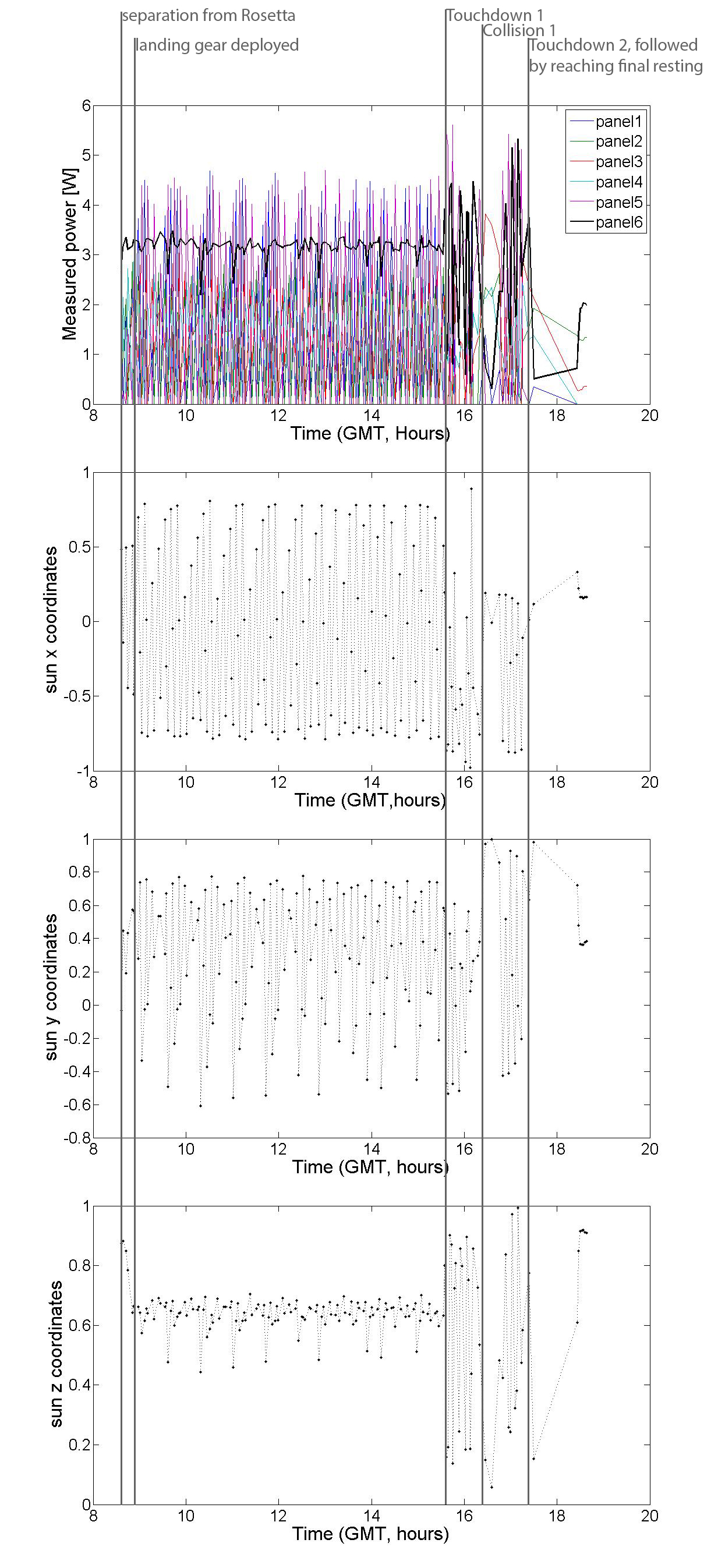}
\caption{Top: in-situ measured power profiles of the six solar panels. Bottom: Reconstructed components of $\vec{n}$}
\label{fig:measured}
\end{center}
\end{figure}
\begin{figure}[h]
\begin{center}
\includegraphics[width=14 cm] {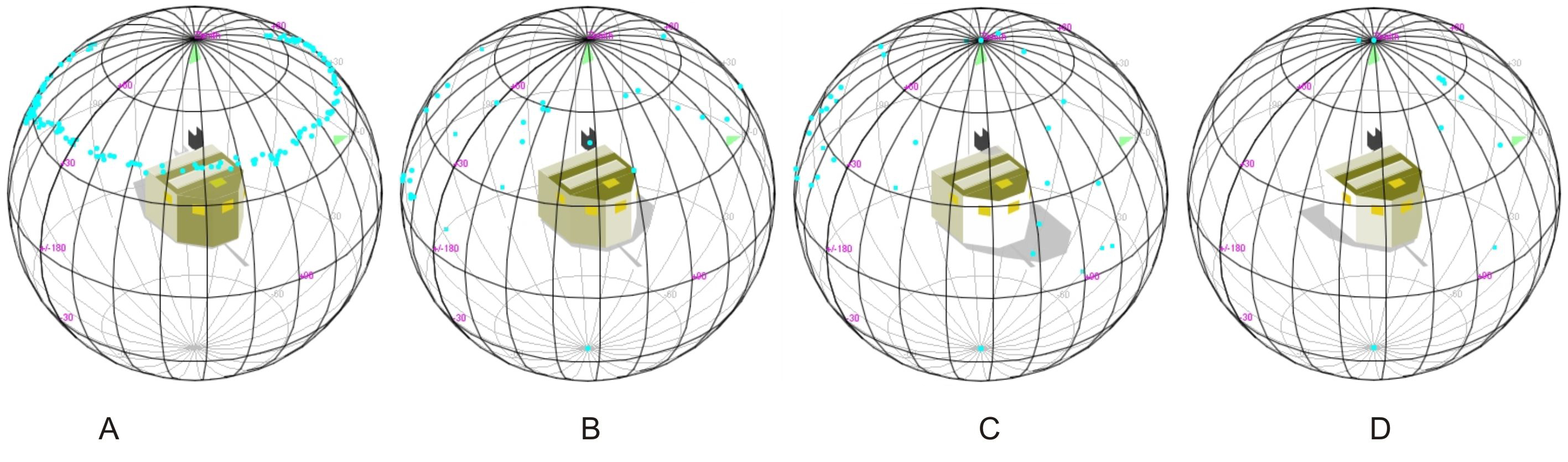}
\caption{Reconstructed directions of the Sun during descent (A), between TD1 and C1 (B), between C1 and TD2 (C) and after TD2 (D).}
\label{fig:trajectories}
\end{center}
\end{figure}

The graphs of reconstructed directions of the  Sun are depicted in Fig. \ref{fig:measured}, from lander separation through the multiple bouncing events over the comet until after reaching a still-stand and the final parking site. Adjacent points of the diagrams are connected by a dotted line for better visibility. Points where the reconstruction was unsuccesful were filtered out upon producing these graphs. Events of the landing process are marked in the figure. The changes of the rotational dynamical at the events are striking. We also show reconstructed directions of the Sun in Fig. \ref{fig:trajectories} during different phases of the motion.

\subsection{The mechanical model of Philae and equations of motion}


The body of the lander is modelled as two rigid bodies: one representing the gyroscope, and another one the main body of the spacecraft (Fig. \ref{fig:geometry}). We use a reference frame x-y-z (from now on: lander frame) attached to the main body. The gyroscope can rotate freely around a shaft parallel to the $z$ axis fixed to the main body. The direction of the shaft of rotation is represented by a unit vector $\vec{u}_Z$. 

Let $\vec{\omega}(t)$ denote the angular velocity of the main body relative to an inertial reference frame (solar frame). Furthermore, let the angular velocity of the gyroscope relative to the main body be expressed as $w_{rot}(t)\vec{u}_Z$. During descent, $w_{rot}(t)$ was kept at a constant level $\omega_{rot0}\approx 865 \ \frac {Rad}{s}$ by a motor, whereas after TD1, the gyroscope was running freely and gradually lost its angular velocity due to a torque $M_f\vec{u}_Z$ caused by internal friction. 

During contact-free motion, the rotational dynamics of this system is not affected by significant external torques. The angular accelerations of the two parts of the lander as functions of $M_f(t)$ can be determined from the conservation of angular momentum 
\begin{align}
\frac{d}{dt}\vec{\omega}=-\left(\matr{\theta}^0_{0}-\vec{u}_Z\vec{u}_Z^T \matr{\theta}^2_2 \right)^{-1}\left(\vec{\omega} \times \matr{\theta}^0_{0} \vec{\omega}+\vec{u}_Z\ M_f+\vec{\omega}\times\matr{\theta}^2_2 \vec{u}_Z\omega_{rot}\right) \label{eq:domega}\\
\frac{ d }{dt}\omega_{rot}=\vec{u}_Z^T(\matr{\theta}^2_2)^{-1}\left(\vec{u}_Z\ M_f-\vec{u}_Z\vec{u}_Z^T\matr{\theta}^2_2  \frac{d}{dt} \vec{\omega}\right)\label{eq:domegarot}
\end{align}
as derived in the Appendix. The arguments $(t)$ of $\vec{\omega}$, $\omega_{rot}$ and $M_f$ have been dropped in the equations for brevity; $\matr{\theta}^2_2$ refers to the mass moment of inertia tensor of the gyroscope with respect to its center of mass, and $\matr{\theta}^0_0$ denotes the mass moment of inertia tensor of the lander and the gyroscope with respect to their joint center of mass. Estimated values of the moments of inertia in lander frame are summarized in Table \ref{tab:inertia}. 

\begin{figure}[h]
\begin{center}
\includegraphics[width=7
 cm]{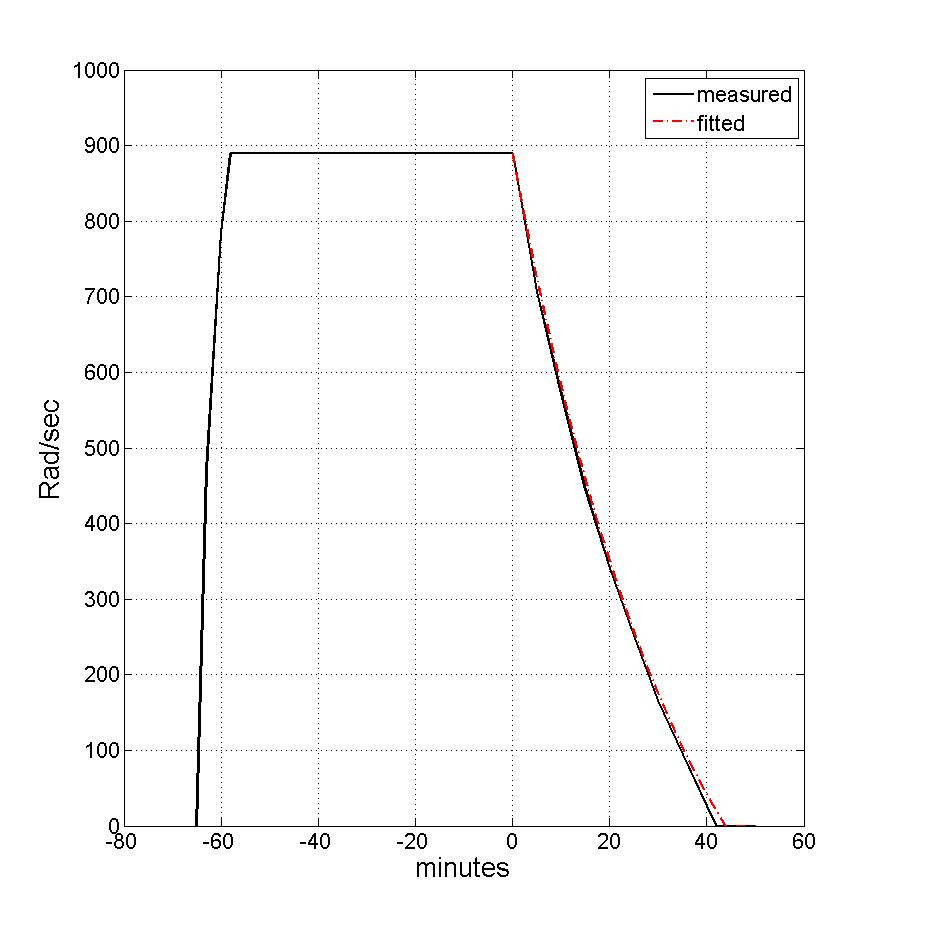}
\caption{Solid line: angular velocity of the gyroscope measured  during an in-flight test for fly-wheel characterisation and received as part of Philae telemetry data. Dash-dotted line: fitted exponential function  }
\label{fig:gyroscope}
\end{center}
\end{figure}

The frictional torque $M_f(t)$ at the shaft of the gyroscope can be deduced from measured $w_{rot}(t)$ function during an in-flight test (Fig. \ref{fig:gyroscope}). We have fitted an exponential function $\bar{w}(t)=c_1 e^{c_2t}+c_3$ to the decaying part of the $w_{rot}(t)$ diagram, yielding $c_1=1271$, $c_2=-0.0004643$, $c_3=-388.9$ (dashed line in Fig. \ref{fig:gyroscope}). Exponential functions of this form satisfy the identity
$$
\frac{d}{dt}\bar{w}(t)=-\left(c_2c_3-c_2\bar{w}(t)\right)
$$ 
From this formula, we deduce an empirical law of sliding friction:
\begin{equation}
M_f= -\sign(\omega_{rot}) \cdot (\mu_0+\mu_1\abs{w_{rot}}) \quad if\quad \omega_{rot}\neq 0 
\label{eq:Mslip}
\end{equation}
with $\mu_0=c_2c_3\theta^2_{2,zz}=1.0\cdot10^{-3}$ and $\mu_1=-c_2\theta^2_{2,zz}=2.6811 \cdot 10^{-6}$
where $\theta^2_{2,zz}$ is the principal moment of inertia of the gyroscope about the $\vec{u}_z$ axis (see Table \ref{tab:inertia}).

In the spirit of Coulomb's classical friction law, we assume the following extension for the case of an immobile gyroscope:
\begin{equation}
\left\{
\begin{array}{cccc}
\quad \dot\omega_{rot}=0 & and & |M_f|< \mu_0 & or \\
M_f=-\mu_0 &  and &  \dot\omega_{rot}>0 &  or \\
M_f=\mu_0& and & \dot\omega_{rot}<0 &   
\end{array}
\right\} \quad if \quad \omega_{rot}=0
\label{eq:Mstick}
\end{equation}
where dot represents derivation with respect to time. The implicit formula \eqref{eq:Mstick} always determines $M_f$ and $\dot{\omega}_{rot}$ uniquely, i.e. exactly one of the 3 cases yields a consistent solution.



%
%
\begin{table}[h]
  \begin{center}
    \caption{Estimated mechanical properties of Philae ($\Theta$ in $kgm^2$)}
    \label{tab:inertia}
    \begin{tabular}{|c|c|c|c|c|c|c|c|}
    \hline
      m & $\theta^0_{0,xx}$ & $\theta^0_{0,xy}$ & $\theta^0_{0,xz}$ & $\theta^0_{0,yy}$& $\theta^0_{0,yz}$& $\theta^0_{0,zz}$& $\theta^2_{2,zz}$ 
      \\
      \hline
      97.63 kg & 13.905 & -0.352  & -0.015  & 12.603  &0.007 & 16.523 & 0.0058 
      \\
      \hline
    \end{tabular}
  \end{center}
\end{table}

\begin{figure}[h!]
\begin{center}
\includegraphics[scale=0.15]{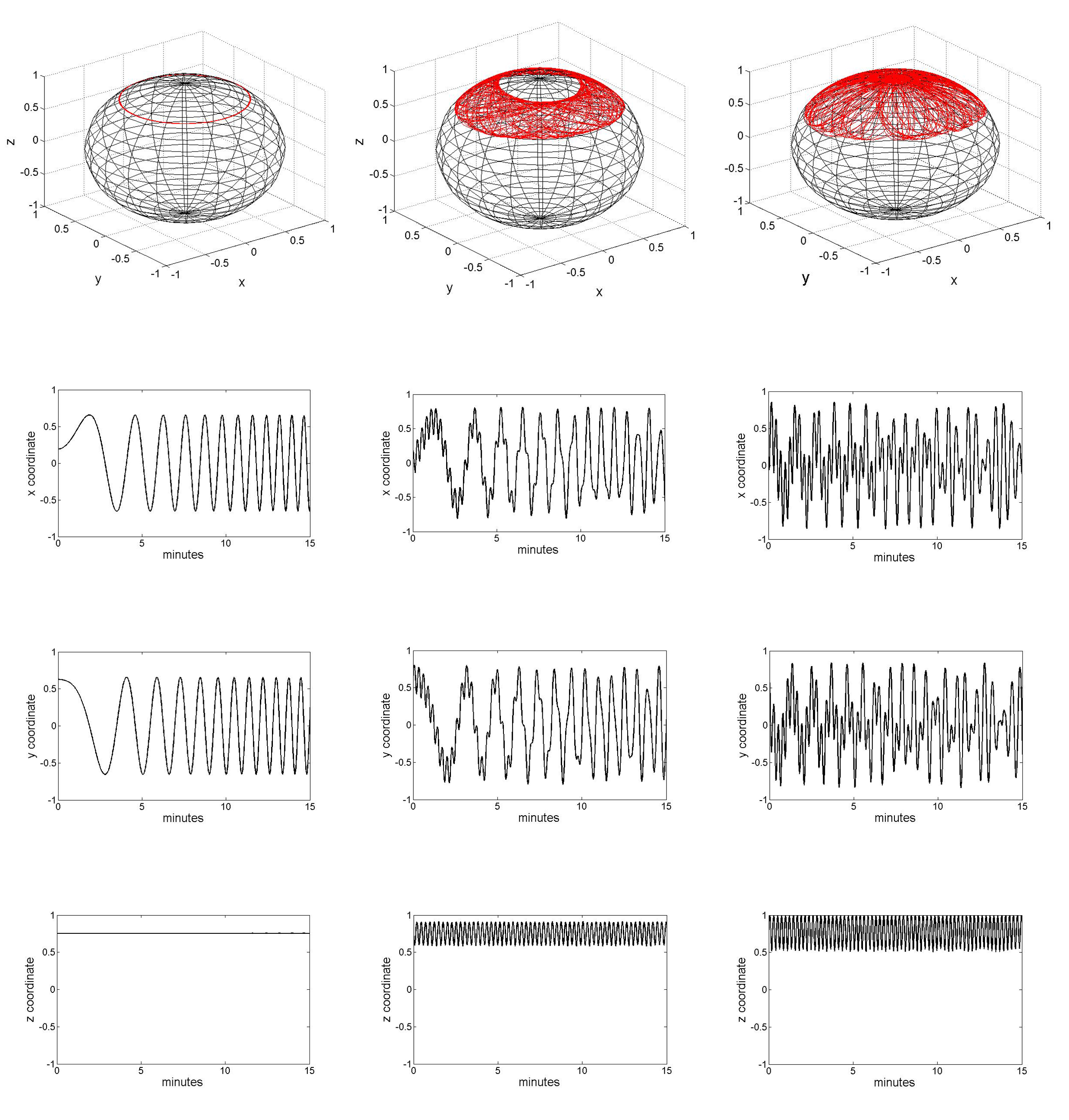}

\caption{Trajectories of the endpoint of $\vec{r}(t)$ (top) and the three components of $\vec{r}(t)$  during simulated motion of the sun in lander frame. The length of each simulations is 46 minutes. Initial values are $\vec{\omega}=[0;0;0]$ (left); $\vec{\omega}=[0.1;0;0]$ (middle); and $\vec{\omega}=[-0.1;0.2;0]$ (right),  with $r=[0.1959;0.5690;0.7986]$ and $\omega_{rot}=865$ $\frac{Rad}{s}$ in all cases. }
\label{fig:simulation}
\end{center}
\end{figure}

\begin{table}[|c|c|c|c|]
\centering
\caption{Coordinates of the Legs}
\label{tab:legs}
\begin{tabular}{|c|c|c|c|}
\hline
 & $\vec{l}_1$ & $\vec{l}_2$ & $\vec{l}_3$      
\\\hline
x & 1.4780  & -0.7784 & -0.7784 					\\\hline
y & 0.0118 & 1.3775 & -1.3539 					\\\hline
z & -0.4078 & -0.4078 & -0.4078            		 	\\\hline

\end{tabular}
\end{table}

The present work focuses on measured and simulated trajectories of the sun in lander frame. We continue using symbol $\vec{n}$ for directions reconstructed from measurements, and introduce the unit vector $\vec{r}(t)=[r_x(t) r_y(t) r_z(t)]$ denoting the time-dependent direction of the Sun in simulations expressed in lander frame. Owing to the great distance of the lander from the Sun, the direction of the sun from Philae can be considered constant in solar frame. In lander frame, the sun appears to rotate according to the kinematic relation
%
%
%
%
\begin{align}
\dot{\vec{r}}(t)=-\vec{\omega}(t) \times \vec{r}(t).
\label{eq:suntrajectory}
\end{align}

\begin{figure}[h!]
\begin{center}
\includegraphics[scale=0.15]{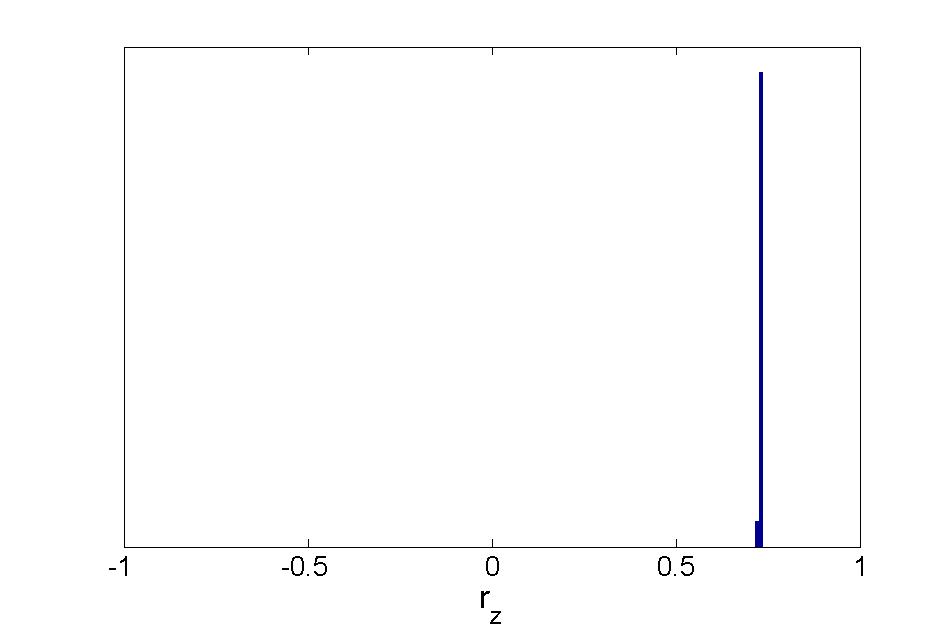}
\includegraphics[scale=0.15]{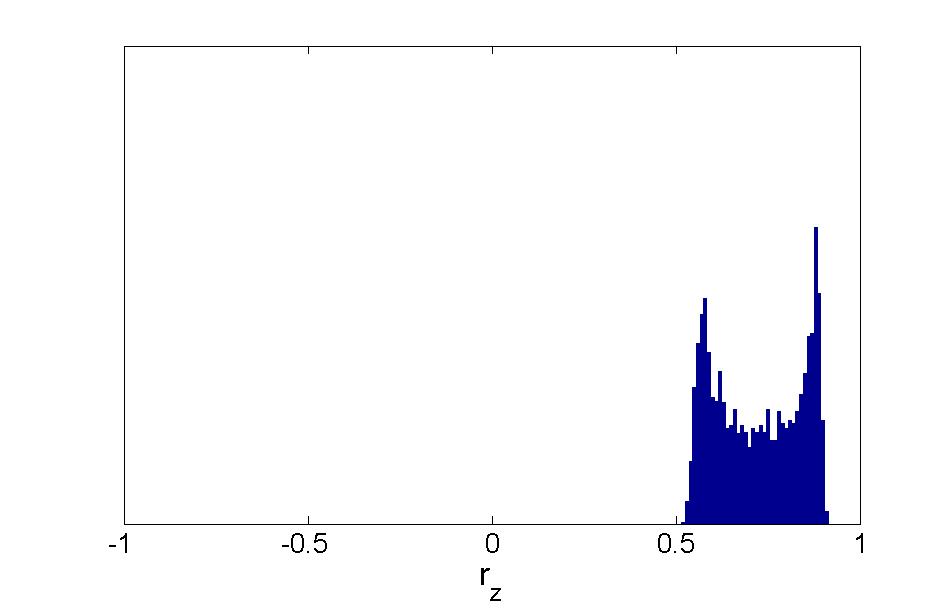}
\includegraphics[scale=0.15]{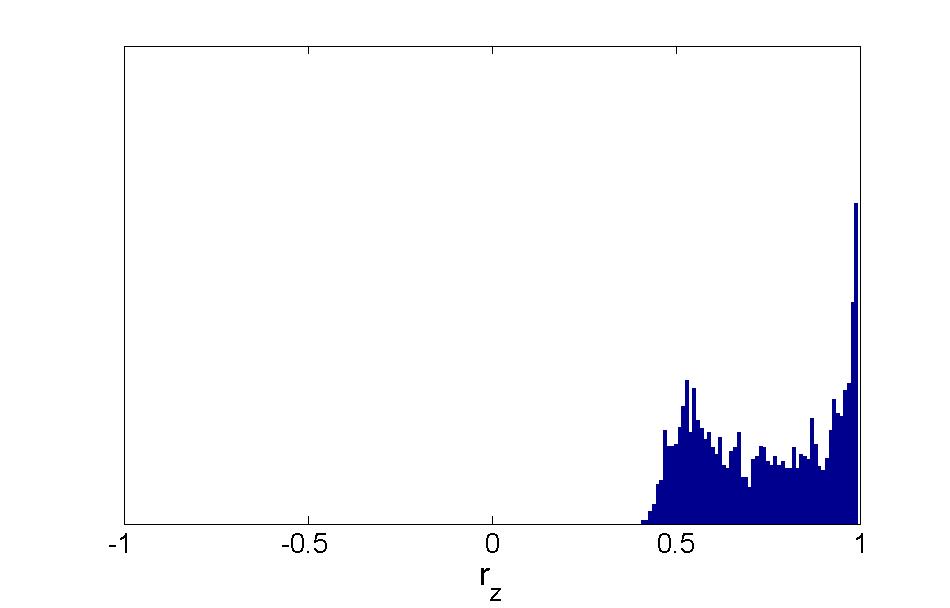}
\caption{Histograms of $\vec{r}_z$ in the simulations of Fig. \ref{fig:simulation}}
\label{fig:histogramms}
\end{center}
\end{figure}

Equations \eqref{eq:domega}--\eqref{eq:suntrajectory} together form a system of ordinary differential equations, which has been simulated numerically in MatLab environment using the ODE solver \textit{ode113} forward and also backward in time from given initial conditions. 
%
Several examples of forward simulations are depicted in Fig. \ref{fig:simulation}. In the first case (Fig. \ref{fig:simulation}.A), the initial value of $\vec{\omega}(t)$ is parallel to the $z$ axis of the lander frame. The axis of rotation is stabilized by the gyroscope, and thus the sun goes around in circles around the $z$ axis. Accordingly, $r_z$  remains constant. Internal friction decelerates the gyroscope, and accelerates the rotation of the lander around the $z$ axis. This tendency is reflected by the shape of the oscillation in the diagrams of $r_x(t)$ and $r_y(t)$ and it is also confirmed by ROMAP measurements \cite{ESAblog2}. In the other two cases, the initial values of $\vec{\omega}$ have non-zero $x$ and $y$ components, which initiates rotation accompanied by precession. Accordingly, the trajectory covers a ring- or cap-shaped part (or the whole) of the sphere. The gradual speed-up of rotation around the $z$ axis is again visible from the $r_x(t)$ and $r_y(t)$ diagrams. A closer inspection of the $r_z(t)$ diagrams reveals yet another important property. The range in which $r_z$ oscillates depends on the initial conditions, but it remains roughly the same during motion in spite of the gradually slowing gyroscope. We will exploit this fact later when we analyse histograms of $r_z$ (Fig. \ref{fig:histogramms}). These histograms reflect the degree of precession of the $z$ axis of the lander. At the same time, the histograms of $r_x$, $r_y$ bear little information about the actual motion of the lander, since their shapes are determined largely by the fast rotation about the $z$ axis initiated by the gyroscope.

Our analysis includes simulations of the collision C1 of the lander to the comet surface, for which the following simplified model has been used. Let $\vec{F}$ denote the impulsive force transferred by the ground to the lander at the point of collision having position vactor $\vec{l_i}$ in lander frame. We will assume that the point of collision is the endpoint of one of the legs, hence the notation $l_i$ where $i=1,2$ or $3$. Let $\delta\vec{\omega}$ and $\delta\vec{v}$ denote the jumps in its angular velocity and in the velocity of its center of mass. 
The impact process is modelled by an instantaneous event, during which the contact force is large, and the effect all forces other than $\vec{F}$ is negligible. It is assumed that the effect of $M_f$ is also negligible. With this assumptions, we can write a discret-time analogue of \eqref{eq:domega} 
\begin{align}
\delta\vec{\omega}=\left(\matr{\Theta}^0_{0}-\vec{u}_Z\vec{u}_Z^T \matr{\Theta}^2_2 \right)^{-1}\left(\vec{l}_i\times\vec{F}\right).
\label{eq:impmoment}
\end{align}
in which the cross-product terms are omitted as their effect is negligible during the infinitesimally short duration of the impact. Similarly the conservation of linear momentum yields: 
\begin{align}
\delta\vec{v}=\frac{1}{m}\vec{F},
\label{eq:imp}
\end{align}
where $m$ is the mass of the lander. The impulse $\vec{F}$ is determined from the assumption of an inelastic, sticking impact, wherein the velocity of the contact point $\vec{l}_i$ becomes $\vec{0}$:

\begin{align}
(\vec{\omega}+\delta\vec{\omega} ) \times \vec{l}_i
+\vec{v}+\delta\vec{v} 
= 0
\label{eq:collisioncriteria}
\end{align}
Eq. \eqref{eq:impmoment} -- \eqref{eq:collisioncriteria} form a system of linear equations, which can be solved for $\vec{F}, \ \delta\vec{\omega}$ and $\delta\vec{v}$.
The last step of resolving the impact is to determine the instantaneous jump $\delta\omega_{rot}$ of the angular velocity of the gyroscope  from the discrete-time analogue of \eqref{eq:domegarot}:
\begin{equation}
\delta\omega_{rot}=-\vec{u}_Z^T\delta\vec{\omega}
\label{eq:deltaomegarot}
\end{equation}
by which the post-impact dynamic state of the lander becomes fully known.



\subsection{Comparison of simulated and measured sun locations}

In this subsection, we define two metrics of the distance between the simulation results, and the array of measured sun locations. These metrics will be minimized in order to identify those initial conditions of the simulation, which  provide the best fit to measured data.


\subsubsection*{Direct fitting}
The natural goal of data fitting is to look for a simulated trajectory going through the measured locations of the sun at the times of the measurements. This approach corresponds to minimization of the error function
\begin{align}
E=\frac{1}{2}\text{mean} (1-\vec{r}(t_i)^T\vec{n}_i),
\end{align}
where the unit vector $\vec{n}_i$ denotes the $i$-th reconstructed sun location at time $t_i$ and $\vec{r}(t_i)$ is the simulated sun location at the same time. The $E$ metric becomes zero in the case of a perfect fit, and its theoretical maximum value is 1. We make use of the $E$ metric for trajectory reconstruction before TD1. During this period, the motion of the lander is a precession-free rotation about its $z$ axis and angular velocities can be reconstructed accurately using trajectory fitting (see Sec. 3.1).

\begin{figure}
\begin{center}
\includegraphics[scale=0.2]{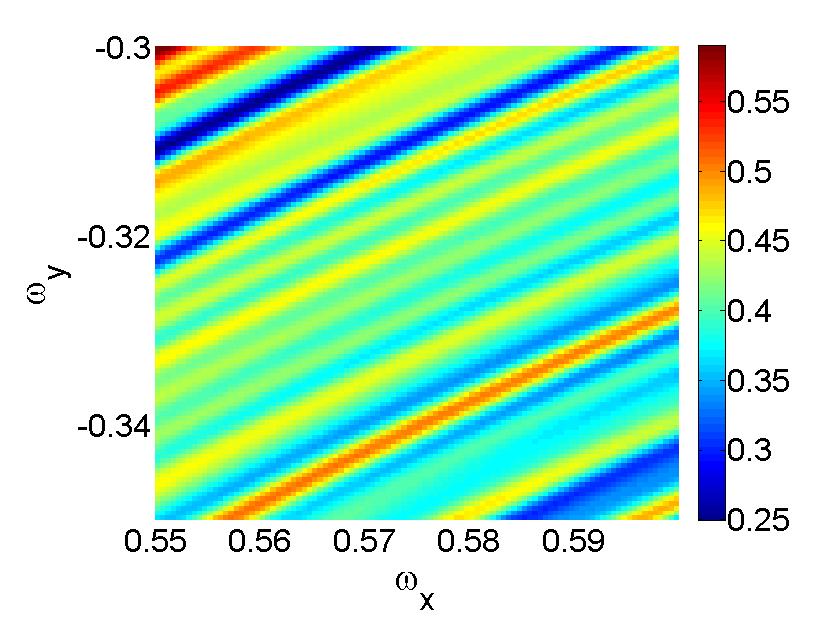}
\includegraphics[scale=0.2]{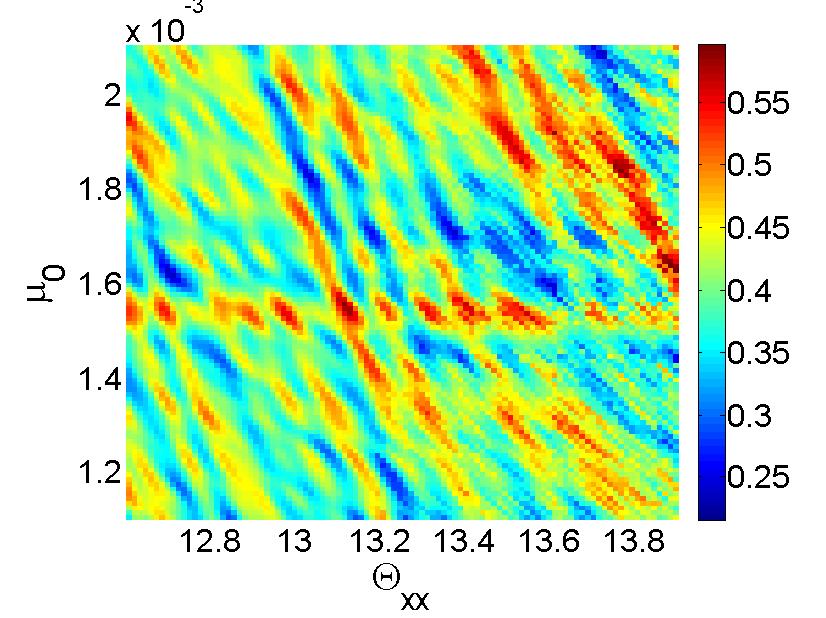}
\caption{A: Evaluation of $E$ in a small region with initial condition $\omega_x\in [0.55 \ ...0.6],\omega_y\in [-0.33 \ ...-0.3], \omega_z=0.3$. B: Evaluation of $E$ with initial condition $\vec{\omega}=[0.55;-0.35;0.3]$ as a function of $\mu_0\in[1.1 \  ...2.1]\cdot 10^{-3}$ and $\theta_{xx}\in [12.603 \ ...13.905]$. Other parameters are given by table \ref{tab:inertia}.}
\label{fig:perturbations}
\end{center}
\end{figure}

The metric $E$ has a fundamental drawback when applied to trajectory reconstruction of combined rotational and precessional motion after TD1: it appears extremely sensitive to variations of system parameters (mass moments of inertia, friction law) and initial values of motion ($\vec{\omega}$, $\omega_{rot}$)  The sensitivity is a consequence of the low sampling rates (one measurement per 2 minutes) and the long time window (40 minutes between TD1 and the next collision), because small perturbations of model parameters change the frequency of rotation and precession to some extent, which leads to the accumulation of large phase shifts during a long time window. This effect causes large variations in $E$. The sensitivity is illustrated by the complex shape of the diagram of $E$ versus parameters and initial condition in Fig. \ref{fig:perturbations}. The data of the figure are based on simulated trajectories of length 45 min. (the time from the first data point after TD1 to C1), in which the initial direction of the Sun was $\vec{r}=[0.1918;0.6274;0.7071]$ (the first measured direction of the Sun after TD1) , the initial angular velocity of the gyroscope was $\omega_{rot}=865\frac{Rad}{s}$ (estimated value at first data point shortly after TD1). For other parameter values, see figure caption. 

The reasons highlighted above make this metric unsuitable for the reconstruction of complex trajectories with precession after TD1. For this purpose, a second metric is developed.

\subsubsection*{Fitting the probability distribution of $z$ coordinates}
The second metric is based on comparing  probability density functions (PDF) of the $r_z$ coordinate of the Sun with the histogram of measured values $n_z$.  We have seen that the PDFs of the $r_x$ and $r_y$  coordinates bear little information as they are determined largely by the fast rotation around the $z$ axis. Nevertheless the PDF of the $r_z$ coordinate reflects the precessional component of the motion, which is in the focus of our interest. Comparing  PDFs instead of the $r_z(t)$ functions themselves has the advantage of being less sensitive to small errors in model parameters and initial conditions. As we have pointed out earlier, the root of the sensitivity is at the accumulating phase-shifts of rotational and precessional motion, and the PDFs of the location of the Sun are not affected by such phase shifts. We have also found in Sec. 2.3 that the amplitude of precession (reflected by the amplitude of oscillation of $r_z(t)$) does not change significantly as the gyroscope slows down. Hence an inaccurate model of friction between the main body of the lander and the gyroscope does not alter the PDF of $r_z$ significantly.

%
%

We use the Earth Mover's Distance, or EMD \cite{EMD1,EMD2,EMD3} (also known as Wasserstein-metric or Mallows-distance) as a metric of distance between measured and simulated distributions. In the case of one dimensional distributions, the EMD can be calculated as

\begin{align}
EMD=\frac{1}{2} \int_{-1}^1 \! \abs{C_r(z)-C_n(z)} \, \mathrm{d}z,
\label{eq:EMD}
\end{align}
where $C_r$ and $C_n$ stand for the cumulative distribution of reconstructed and simulated z coordinates of the sun respectively. The theoretical minimum value of the EMD is 0 (if the two distributions are identical), while the theoretical maximum is 1. 

The equation \eqref{eq:EMD} takes continuous distributions as inputs. Nevertheless the measured data are discrete, and the simulated data are also discretized using a sufficiently dense sampling of trajectories over time. Discrete distributions can be represented by the dirac delta function, which enables us to evaluate \eqref{eq:EMD}.

In order to obtain accurate results, our EMD-based metric takes into account the fact that not all directions of the sun can be reconstructed from solar panel power data. Hence, the probability distribution functions of the simulation are generated only from those portions of the simulated trajectories, for which reconstruction is possible. 

\begin{figure}
\begin{center}
\includegraphics[scale=0.2]{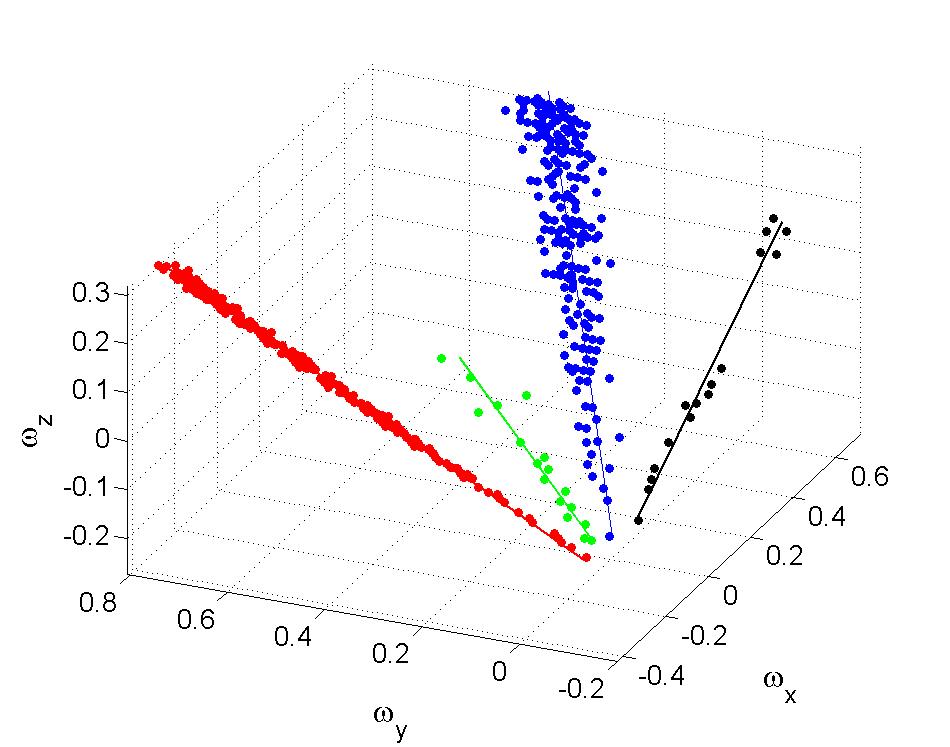}
\caption{Points in the velocity-space satisfying \ref{eq:EMDcrit} with fitted straight lines }
\label{fig:EMD}
\end{center}
\end{figure}

The EMD was determined in a series of numerical simulations. The simulation is assumed to start at the time of the first solar panel power measurement, for which the reconstructed direction of the Sun is $\vec{n}=[0.1959;0.5690;0.7986]$. The components $\omega_x,$ and $\omega_y$ of the initial angular velocities were systematically varied within the interval $[-0.5...1]$, and $\omega_z$ in $[-0.3...0.3]$ 
with a step size of $0.015 [rad/s]$ in all directions
. For other initial conditions and parameter values, see previous section. The minimum value  of the EMD is $0.0326$, which indicates a reasonably good fit. Nevertheless the EMD does not have a characteristic minimum point, but it takes values close to its minimum along four lines.  Fig. \ref{fig:EMD} shows points in $\mathbb{R}^3$ where

\begin{align}
\label{eq:EMDcrit}
EMD(\vec{\omega}) < 1.1 \cdot EMD_{min}.
\end{align}

Four straight lines have been fitted to these points by minimizing the sum of Euclidian distances from the line to the points. This gave the lines (with $\lambda_i\in \mathbb{R}$):

\begin{align}
L1:&=[0.5148;0.3102;0.1391]+\lambda_1[0.6797;0.4305;0.5939]\\
L2:&=[-0.2116;0.5308;0.1185]+\lambda_2[-0.2908;0.7447;0.6007]\\
L3:&=[-0.2141;0.0559;0.0000]+\lambda_3[-0.5761;0.1806;0.7972]\\
L4:&=[0.2247;-0.1044;0.0366]+\lambda_4[0.5398;-0.2442;0.8056]
\end{align}

We believe that the existence of minimum lines can be explained by the fact that typical distributions of $r_z$ have similar shapes to one another with 2 distinct peaks. Such distributions can be characterized by two scalar parameters, such as the locations of the peaks, or the mean and the standard deviation of the distribution. In this situation, a good fit of the simulated distribution requires that two constraints are satisfied, which results in a one-dimensional solution set in 3D space.
Having a one-dimensional set of optima has fundamental consequences for our analysis based on fitting statistical distributions. We are not able to identifiy the trajectory of the lander, nevertheless the analysis provides a one-dimensional set of solution candidates. This is why the reconstruction presented in this paper remains partial.




\section{Inferences for the motion of Philae}

\subsection{Before TD1} \label{sec:3.1}
The motion of the lander before TD1 is well understood: the lander performed slow rotation around the $z$ axis without significant precession. The gyroscope was kept at roughly (but not perfectly) constant rate of rotation, i.e. the internal friction was compensated by a motor. Our motivation for comparing simulations and measurements during this period is to verify that the Sun reconstruction method of Sec. 2.1 yields fairly accurate results.

The measured $n_{z}$ values tend to be near $0.8$ during the first few measurements, and around $0.60$ thereafter (Fig. \ref{fig:measured}). The constant $n_z$ value during most of the time is the consequence of pure rotation about the z axis, whereas initial variations of the measured $z$ coordinates is caused by the deployment of the legs shortly after the release of Philae. In the closed position of the legs, some of the solar panels are partially shadowed, which alters the results of reconstruction. High values of the $\Delta$ error function also indicate the disturbing effect of the legs during the initial period of motion. Because of this unmodeled effect, we concentrate on the motion after the deployment of the legs. The timespan of the studied motion was about six and a half hours
ending with measurement number 143.

\begin{figure}
\begin{center}
\includegraphics[scale=0.25]{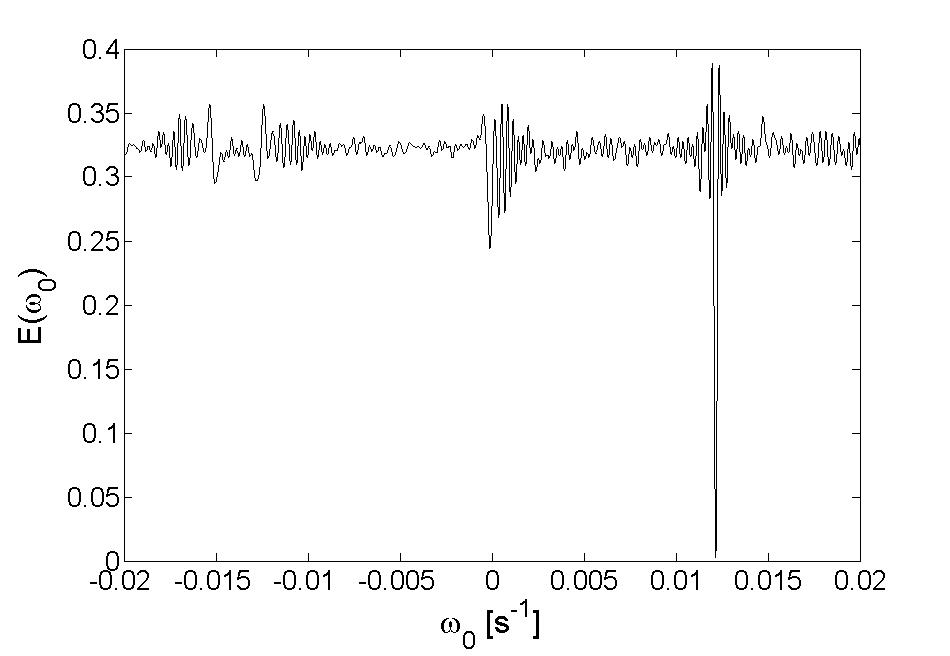}
\caption{Evaluation of $E(\bar{\omega}(t))$, as function of $\omega_0$}
\label{fig:E_abra}
\end{center}
\end{figure}

The angular velocity of the lander is assumed to be $\vec{\omega(t)}=[0\;0\;\bar{\omega}(t)]$ where the nearly constant function $\bar{\omega(t)}$ is approximated by the polynomial $\bar{\omega}(t)=\omega_3 t^3+\omega_2 t^2 +\omega_1t+\omega_0$ ($t$ measured in units of $s$ and $\bar{\omega}(t)$ in $s^{-1}$). The time $t=0$ corresponds to the time of the first measurement after deployment of the legs (08:50:30 time UTC, measurement number 10). The coefficients $\omega_i$ have been optimized numerically for a minimum value of metric $E$. A well-defined, unique optimum has been found at $\omega_0=0.01216$, $\omega_1= 5.161 \cdot 10^{-8}$, $\omega_2=-3.189 \cdot 10^{-12}$, $\omega_3=3.013 \cdot 10^{-17}$. The corresponding value of $E=0.0028$ indicates a very good fit, as illustrated by Fig. \ref{fig:E_abra}. The reconstructed angular velocity varies in the range of $0.01216...0.01224 s^{-1}$ during the descent phase, which matches well with the value $0.0123s^{-1}$ estimated from ROMAP sensor on-board Philae \cite{ESAblog2}.


\begin{figure}
\begin{center}
\includegraphics[scale=0.25]{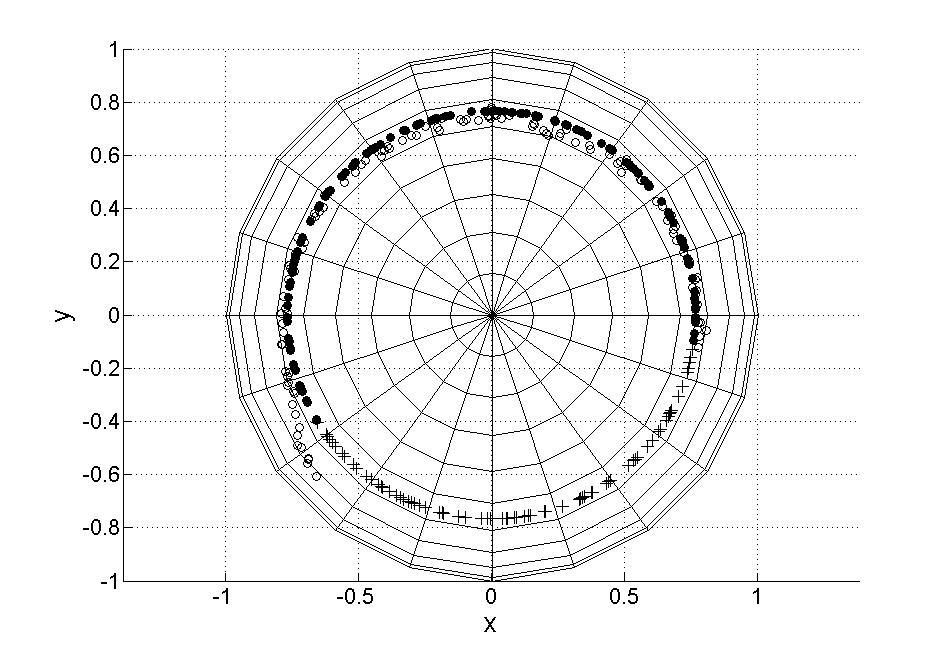}
\caption{Comparison of simulated ($\bullet$) and measured ($\circ$) sun locations after trajectory fitting. The points shown with $+$ are locations of the Sun, which are predicted by numerical simulation to lie behind the lander, where solar power measurements do not allow the reconstruction.}
\label{fig:comparison}
\end{center}
\end{figure}

\subsection{Between TD1 and C1} \label{sec:3.2}
The ground reaction forces of TD1 increased the angular velocity of the lander abruptly. Even though the dampers in the legs absorbed over 90\% of the kinetic energy, the motion did not stop completely. In particular, a faster rotation accompanied by precession emerged. 

The two most interesting questions about the lander's motion are the amplitude of precession and the magnitude of the angular velocity. The first one can be read directly from the measurements and the results of parameter fitting: these show that the $z$ coordinate of the Sun oscillated roughly in the range of $(10^o...80^o)$. The size of this interval corresponds to the amplitude of the oscillation of the $z$ axis of the lander in global frame. The relatively small interval suggests that the lander had its legs pointing towards the local comet surface after TD1 (provided the surface was not very rough in that area). This is important as it suggests that the lander may have hit an obstacle with one of its leg (rather than with its main body) upon event C1.  

\begin{figure}
\begin{center}
\includegraphics[scale=0.2]{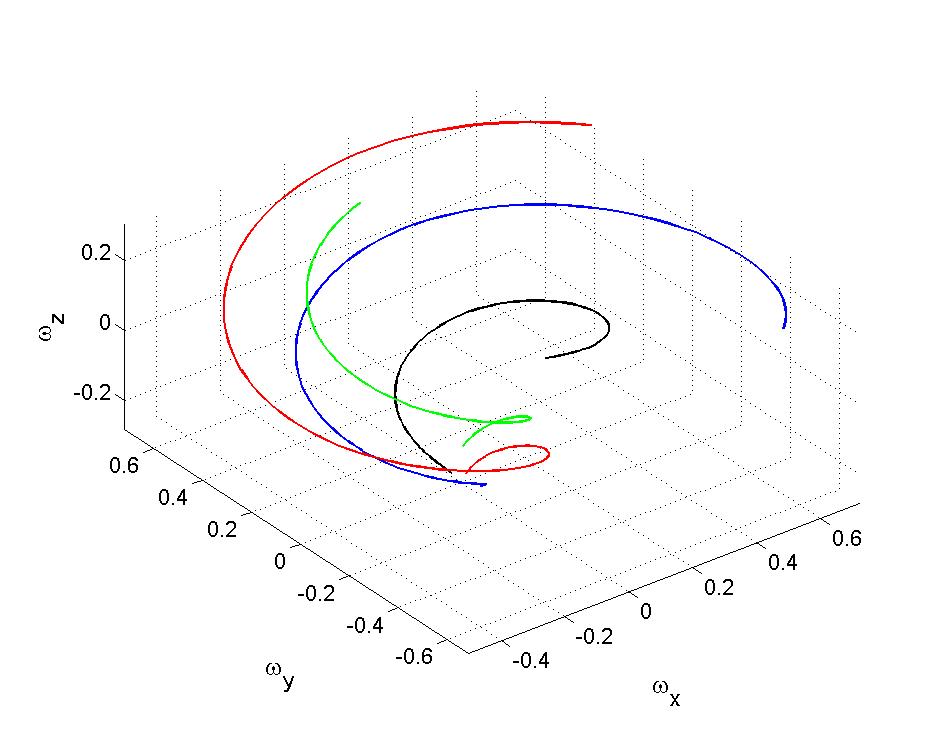}

\caption{Possible angular velocities at the time of TD1 in ($\frac{Rad}{s}$)}
\label{fig:backwards}
\end{center}
\end{figure}

To get a picture about the angular velocity immediately after TD1, we made use of the one-dimensional set of candidate values for the angular velocity at the time of the first solar panel measurement after TD1, as described in Sec. 2.3. These angular velocities have been used as initial conditions of numerical simulation backward in time until the time of TD1. The result (Fig. \ref{fig:backwards}) is another set of four curves. An interesting property of these curves is that they are distant from the origin. In particular, our analysis predicts that the angular velocity after TD1 was at least $0.1816 \ s^{-1}$.  

\subsection{After C1} \label{sec:3.3}
\begin{figure}[h]
\begin{center}
\includegraphics[scale=0.2]{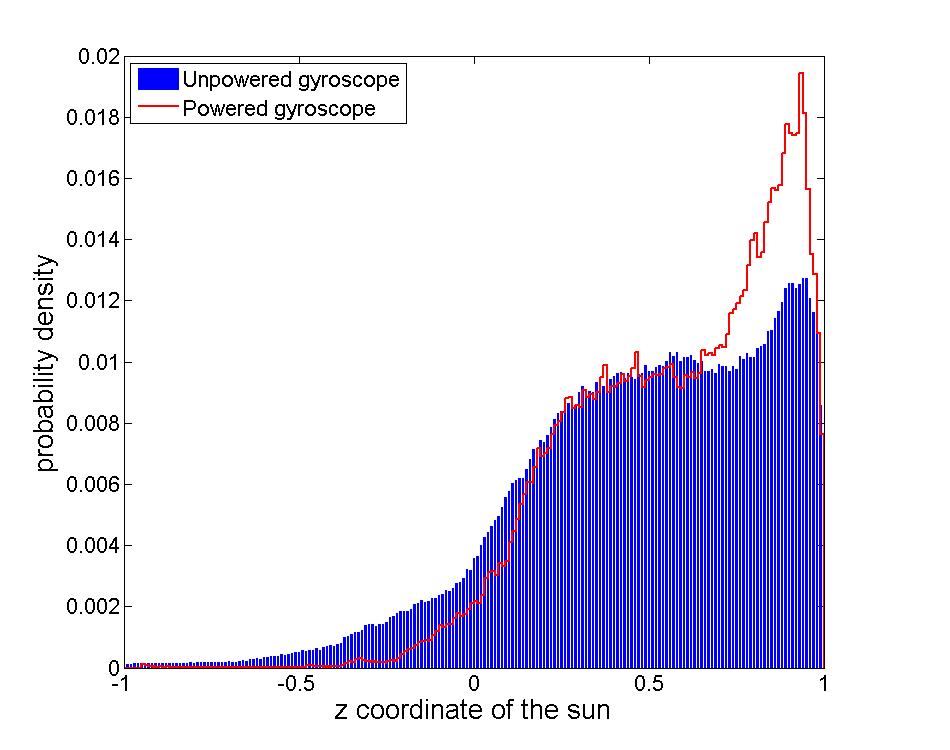}
\caption{Normalized histogram of 400 random simulations between C1 and TD2. The blue bars show the results with the gyroscope switched off, while the red line stands for the case of it having a constant speed. The area under both distributions is 1.}
\label{fig:afterC1}
\end{center}
\end{figure}

The collision event C1 again changed the motion of the lander abruptly. For the time interval between C1 and TD2, there are only 17 available solar panel measurments. These measurements show that the amplitude of oscillation of the $z$ coordinate of the Sun increased and it often took negative values, for which the direction of the Sun is not reconstructible from measurements. 
 Nevertheless the amount of measurements is clearly insufficient for parameter fitting. Hence, we rely on numerical simulations to infer what may have happened. To this end, 400 random candidate angular velocities from Fig. \ref{fig:EMD} have been picked and simulated forward until C1. The simulation yields possible values of lander attitude and angular velocity at the time of C1. An estimated value $v=0.2355$ $\frac{m}{s}$ of the velocity of translational motion during this phase of motion was obtained from the fact that Philae travelled about 650 meters along the surface of the comet in 46 minutes \cite{auster15}. The direction of the velocity vector was chosen randomly with the condition that its z coordinate (in lander frame) was negative. It is not known, which leg may have touched the ground during C1, hence one of the three legs was chosen randomly, and the outcome of a collision 
was calculated for all 400 scenarios using the impact model of Sec. 2.2 
After finding post-impact angular velocities, the rotational motion of the lander was simulated forward in time for another 65 minutes (time interval between C1 and TD2). Solid bars in Fig. \ref{fig:afterC1} shows the overall PDF of $r_z$ for all simulations. These result suggest that the $z$ coordinate of the Sun may have reached all parts of the $(-1,1)$ interval. In other words, the direction of the lander $z$ axis probably lost its relative steadiness in solar frame. This was a dangerous situation for the success of landing for two reasons: 
first, any part of the lander (including legs or body) may have hit the ground upon TD2, which posed the solar panels to a risk of injury. Indeed, it was verified after Philae wake-up from hybernation that none of the solar panels was damaged during the multiple touch-down and bouncing phases.

Second, the lander was at the risk of coming to rest upon some of the solar panels, with its legs losing contact. Such an unwanted situation could have blocked communication with Rosetta because the  radio antenna of Philae had a visibility cone with opening angle $120^o$ pointing towards the positive $z$ axis.

\section{Summary of results and perspectives for future missions}
In this paper we used measurements of solar panel power profiles as subsets of Philae’s on-board measured telemetry data packages to reconstruct the direction of the Sun relative to the lander during its descent and landing, and to reconstruct its rotational motion. While the limited amount of measured data did not allow an exact reconstruction, the set of possible motion trajectories was narrowed down to a one-dimensional set. The smallest possible value of the angular velocity immediately after TD1 was found to be as large as $0.18s^{-1}$. The amplitude of precession of the lander was estimated after TD1 and also after the presumable collision C1 with a crater rim. It was demonstrated that there was a significant chance for solar panels to collide with the ground, causing mechanical damage, and of an upside down final resting pose. However it become clear from photographs taken immediately after landing, that resting in an inappropriate pose was luckily avoided by the lander.

One way to identify possible damage of solar panels to check the consistency of power profile data. Due to the geometric arrangement of solar panels, some combinations of the six panels being illuminated are not possible. All power data measured before TD2 are self-consitent, but there are several examples of inconsistency after TD2 in the form of illumination patterns (2-3-6), (2-3) and (2). Interestingly, all of these patterns would become self-consistent with the addition of panel 1, and this is the only way of making all patterns consistent by adding only a single panel. The high values of the $\Delta$ error function of sun reconstruction after TD2 provide another indication of inconsistent data. There are two straightforward explanations for the observed patterns: panel 1 being damaged or being shadowed by the local terrain.
In the view of this observation, it was a risky decision of the Philae control team, to rotate the lander prior to entering the hybernation phase in such a way that panel 1 was exposed to the Sun instead of the smaller panel 2. Nevertheless solar panel power data measured after Philae wake-up from hybernation confirmed the correct operation of panel 1 and the improvement in exposure to solar radiation due to the rotational manoeuvre.      

As a final step of our analysis, possible control strategies of the gyroscope and their effects are discussed briefly. We have seen that the stability of the $z$ axis was completely lost after the C1 event. By this time, the gyroscope stopped due to being turned of upon TD1 
and its angular momentum was transferred to the lander body. The main consequence of the rotation of the lander body was the possibility of a high-velocity impact with the ground: while the translational velocity of the lander was only $v\approx 0.02$ $\frac{m}{s}$, the endpoints of the legs could move by as much as $v+|\vec{l}_i||\vec{\omega}|\approx 0.56$ $\frac{m}{s}$ where $|\vec{\omega}|\approx 0.33$ $s^-1$  is a crude estimation of angular velocity at C1. In order to obtain a picture of the contribution of rotational motion to the destabilization of the lander, the analysis of Sec. \ref{sec:3.3} has been repeated with the assumption that the gyroscope continued to rotate at a constant rate after TD1. The result of these simulations have been added to Fig. \ref{fig:afterC1} as a thin curve. It is noticeable that the $r_z$ coordinate of the sun never decreased below $-0.5$, (and rarely went below $-0.2$) i.e. the degree of destabilization of the lander $z$ axis was less dramatic.
%
%
%
%
%
%
We also examined the consequences of turning off the gyrocope after reaching final resting position. When the gyroscope is finally turned off, our friction model (\ref{eq:Mslip}) predicts a frictional torque of order $M_f\approx 5$ Nm, which could be balanced by frictional forces at the three legs of order $F=/M_f/(3|\vec{l}_i|)\approx1$ N. Given that the weight of the lander in the gravitational field of the comet is only $100 kg \cdot 10^{-3} m/s^2 = 10^{-1} N$, friction would be insufficient to prevent the initiation of sliding motion along the surface. Hence the control strategy chosen by the control team  had clear advantages over the alternative strategy of keeping the gyroscope active until the lander has been confirmed to become immobile. If nothing more, our analysis shows that the design of control strategies of the stabilizing gyroscopes in future missions should take into account the possibility of cerroneous landing scenarios.



\section*{Ackowledgments}
The authors thank Reinhard Roll for useful discussions on the topic and for providing us with approximate values of lander parameters. TB and PLV acknowledge support from
the National Research, Innovation and Development Office of Hungary under grant K104501. AB has been supported by the Hungarian Space Office.

\bibliographystyle{plain}
\bibliography{philaebibliography}





\newpage
\section*{Appendix: derivation of the equations of motion}

We denote velocity, angular velocity and angular acceleration of the lander body relative to the world frame by $\vec{v}$, $\vec{\omega}$ and $\vec{\epsilon}$ respectively. Similarly for the gyroscope we have: $\vec{v}_2$, $\vec{\omega}_2$ and $\vec{\epsilon}_2$. Let us denote the angular velocity of the gyroscope relative to the lander body, around axis $\vec{u}_z$ by $\omega_{rot}$, and its angular acceleration by  $\epsilon_{rot}$. Then,we have 
\begin{equation}\vec{\omega}_2=\vec{\omega}+\vec{u}_Z \omega_{rot}
\label{eq:omega2}
\end{equation}
\begin{equation}
\vec{v_2}=\vec{v}+\vec{\omega} \times \vec{c}_{2}
\label{eq:v2}
\end{equation}
 Differentiation of \eqref{eq:omega2} with respect to time yields
$$\vec{\epsilon}_2=\vec{\epsilon}+\vec{\omega} \times \vec{u}_Z\omega_{rot} +\vec{u}_Z \epsilon_{rot}.$$

Furthermore, let $\matr{\theta}_i^j$ denote the mass moment of inertia matrix of object $i$ with respect to the center of mass of object $j$ where $i,j=1$ means lander body, 2 means gyroscope and 0 means the lander as a whole. Let the contact forces and moments transmitted to the lander body through the bearing of the gyroscope be lumped into a force  $\vec{F}$ acting at $\vec{c}_2$ (center of mass of gyroscope) and a moment $\vec{M}$. Note that the $z$ component of $\vec{M}$ is  identical to the moment $\vec{M_f}$  associated with friction. 
%
%
%
The Euler equations expressing the conservation of angular momentum for the lander body and the gyroscope yield
\begin{align}
\vec{M}+\vec{c}_2\times \vec{F}=\vec{\omega}\times\matr{\theta}_1^0 \vec{\omega}+\matr{\theta}_1^0 \vec{\epsilon}
\label{eq:app1}
\end{align}
%
%
\begin{align}
-\vec{M}=\vec{\omega}_2\times\matr{\theta}_2^2 \vec{\omega_2}+\matr{\theta}_2^2 \vec{\epsilon}_2
\label{eq:app2}
\end{align}

whereas 
Newton's equation for the gyroscope and \eqref{eq:v2} yield
\begin{align}
-\vec{F}=m_2\frac{d}{dt}\vec{v}_2=m_2(\vec{\omega}\times(\vec{\omega}\times\vec{c}_2))+\vec{\epsilon}\times\vec{c}_2)
\label{eq:app3}
\end{align}
because $d\vec{v}/dt=0$ in the absence of external forces. Adding up \eqref{eq:app1} and \eqref{eq:app2}  and replacing $-\vec{F}$ in the expression by the right-hand side of \eqref{eq:app3} gives (after some rearrangement):

\begin{align}
0=\vec{\omega}\times\matr{\theta}_1^0 \vec{\omega}+\matr{\theta}_1^0 \vec{\epsilon}+\vec{\omega}_2\times\matr{\theta}_2^2 \vec{\omega_2}+\matr{\theta}_2^2 \vec{\epsilon}_2+m_2 \vec{c}_2 \times (\vec{\omega}\times(\vec{\omega}\times\vec{c}_2))+\vec{\epsilon}\times\vec{c}_2).
\label{eq:motion1}
\end{align}

Let $\matr{\theta}_2^{20}$ be defined as 
\begin{equation}
\matr{\theta}_2^{20}= \matr{\theta}_2^{0}-\matr{\theta}_2^{2}.
\end{equation}
Then, according to the parallel axis theorem:
\begin{equation}
\matr{\theta}_2^{20}=m_2(\vec{c}_2^T\vec{c}_2I_3-\vec{c}_2\vec{c}_2^T)
\label{eq:parallelaxisthm}
\end{equation}
where $I_3$ is the 3 dimensional identity matrix. Using \eqref{eq:parallelaxisthm}, one obtains 
$$m_2 \vec{c}_2 \times (\vec{\omega}\times(\vec{\omega}\times\vec{c}_2))=\vec{\omega}\times\matr{\theta}_2^{20}\vec{\omega}$$ 
and 
$$m_2 \vec{c}_2 \times (\vec{\epsilon}\times\vec{c}_2)=\matr{\theta}_2^{20}\vec{\epsilon}$$
by which \eqref{eq:motion1} takes the form
\begin{align}
0=\vec{\omega}\times\matr{\theta}_1^0 \vec{\omega}+\matr{\theta}_1^0 \vec{\epsilon}+\vec{\omega}_2\times\matr{\theta}_2^2 \vec{\omega_2}+\matr{\theta}_2^2 \vec{\epsilon}_2+\vec{\omega} \times \matr{\theta}_2^{20}\vec{\omega}+\matr{\theta}_2^{20}\vec{\epsilon}.\label{eq:16}
\end{align}
Individual terms on the right hand-side of (\ref{eq:16})can be expanded (using $\matr{\theta}_2^2 \vec{u}_Z \parallel \vec{u}_Z$) as
\begin{align}
\vec{\omega}_2\times\matr{\theta}_2^2 \vec{\omega}_2=&\vec{\omega}\times\matr{\theta}_2^2 \vec{\omega}+\vec{\omega}\times\matr{\theta}_2^2 \vec{u}_Z\omega_{rot}+\vec{u}_Z\omega_{rot}\times\matr{\theta}_2^2 \vec{\omega}+\vec{u}_Z\omega_{rot}\times\matr{\theta}_2^2 \vec{u}_Z\omega_{rot}\\
=&\vec{\omega}\times\matr{\theta}_2^2 \vec{\omega}-\vec{u}_Z\omega_{rot}\times\matr{\theta}_2^2 \vec{\omega}+\vec{u}_Z\omega_{rot}\times\matr{\theta}_2^2  \vec{\omega} =\vec{\omega}\times\matr{\theta}_2^2 \vec{\omega}\\
\matr{\theta}_2^2 \vec{\epsilon}_2=&\matr{\theta}_2^2 \vec{\epsilon}+\matr{\theta}_2^2 (\vec{\omega}\times\vec{u}_Z\omega_{rot})+\matr{\theta}_2^2 \vec{u}_Z\epsilon_{rot}\\
=&\matr{\theta}_2^2 \vec{\epsilon}+\vec{\omega}\times \matr{\theta}_2^2\vec{u}_Z\omega_{rot}+\matr{\theta}_2^2 \vec{u}_Z\epsilon_{rot}
\end{align}
which transforms (\ref{eq:16}) into
\begin{align}
0=&\vec{\omega}\times\matr{\theta}_1^0 \vec{\omega}+\matr{\theta}_1^0 \vec{\epsilon}+\vec{\omega}\times\matr{\theta}_2^2 \vec{\omega}+\matr{\theta}_2^2 \vec{\epsilon}+\vec{\omega}\times \matr{\theta}_2^2\vec{u}_Z\omega_{rot}+\matr{\theta}_2^2 \vec{u}_Z\epsilon_{rot}+\vec{\omega} \times \matr{\theta}_2^{20}\vec{\omega}+\matr{\theta}_2^{20}\epsilon\\
=&\vec{\omega}\times\matr{\theta}_0^0 \vec{\omega}+\matr{\theta}_0^0 \vec{\epsilon}+\vec{\omega}\times \matr{\theta}_2^2\vec{u}_Z\omega_{rot}+\matr{\theta}_2^2 \vec{u}_Z\epsilon_{rot}
\end{align}
%
The moment of friction can be expressed as the projection of $\vec{M}$ to the $\vec{u}_Z$ axis. With the aid of (\ref{eq:app2}), we obtain
\begin{align}
M_f&=\vec{u}_Z^T(-\vec{M})\\
&=\vec{u}_Z^T(\vec{\omega}_2\times\matr{\theta}_2^2 \vec{\omega_2}+\matr{\theta}_2^2 \vec{\epsilon}_2)\\
&=\vec{u}_Z^T(\vec{\omega}\times\matr{\theta}_2^2 \vec{\omega}+\matr{\theta}_2^2 \vec{\epsilon}+\vec{\omega}\times \matr{\theta}_2^2\vec{u}_Z\omega_{rot}+\matr{\theta}_2^2 \vec{u}_Z\epsilon_{rot}). \label{eq:25}
\end{align} 
Since the gyroscope has cylindrical symmetry about an axis parallel to $\vec{u}_z$, we have $\vec{u}_Z^T(\vec{\omega}\times\matr{\theta}_2^2 \vec{\omega})=0$, and since $\matr{\theta}_2^2 \vec{u}_Z \parallel \vec{u}_Z$, we have

\begin{align}
M_f&=\vec{u}_Z^T\matr{\theta}_2^2 \vec{\epsilon}+\vec{u}_Z^T\matr{\theta}_2^2 \vec{u}_Z\epsilon_{rot}.
\end{align} 
Thereby equations (\ref{eq:app1}),(\ref{eq:app2}) and (\ref{eq:app3}) can be reformulated as 
\begin{align}
0=&\vec{\omega}\times\matr{\theta}_0^0 \vec{\omega}+\matr{\theta}_0^0 \frac{d}{dt}\vec{\omega}+\vec{\omega}\times \matr{\theta}_2^2\vec{u}_Z\omega_{rot}+\matr{\theta}_2^2 \vec{u}_Z\frac{d}{dt}\omega_{rot}\label{eq:app4}\\
0=&\vec{u}_Z^T\matr{\theta}_2^2 \frac{d}{dt}\vec{\omega}+\vec{u}_Z^T\matr{\theta}_2^2 \vec{u}_Z\frac{d}{dt}\omega_{rot}-M_f. \label{eq:app5}
\end{align}
One may multiply (\ref{eq:app5}) with $\vec{u}_Z$ and use the fact that the diagonal matrices $\vec{u}_Z\vec{u}_Z^T$ and $\matr{\Theta}^2_2$ commute, to obtain:
\begin{align}
\matr{\theta}_2^2 \vec{u}_Z\frac{d}{dt}\omega_{rot}=&\vec{u}_ZM_f-\vec{u}_Z\vec{u}_Z^T\matr{\theta}_2^2 \frac{d}{dt}\vec{\omega}.\label{eq:app6}
\end{align}
The unknown accelerations $\frac{ d }{dt}\omega_{rot}$ and $\frac{d}{dt}\vec{\omega}$ can be expressed from equations (\ref{eq:app4}) and (\ref{eq:app6}), yielding the equations of motion (\ref{eq:domega}) and (\ref{eq:domegarot}).




\end{document}